# Symmetry breaking in spin spirals and skyrmions by in-plane and canted magnetic fields

L. Schmidt, J. Hagemeister, P.-J. Hsu, A. Kubetzka, K. von Bergmann and R. Wiesendanger

Department of Physics, University of Hamburg, Jungiusstr. 11, 20355 Hamburg, Germany
lschmidt@physnet.uni-hamburg.de

**Abstract**

The influence of in-plane and canted magnetic fields on spin spirals and skyrmions in atomic bilayer islands of palladium and iron on an Ir(111) substrate is investigated by scanning tunnelling microscopy at low temperatures. It is shown that the spin spiral propagation direction is determined by the island's border which can be explained by equilibrium state calculations on a triangular lattice. By application of in-plane fields, the spin spiral reorientates its propagation direction and becomes distorted, thereby allowing a proof for its cycloidal nature. Furthermore, it is demonstrated that the skyrmions' shape is distorted in canted fields which allows to determine the sense of magnetisation rotation as enforced by the interfacial Dzyaloshinskii-Moriya interaction.

1. Introduction

Ultrathin magnetic films on heavy metal substrates can exhibit complex spin textures due to the competition of Heisenberg exchange interaction, magnetocrystalline anisotropy, and Dzyaloshinskii-Moriya interaction (DMI) [1]. A significant contribution of DMI is related to a strong spin-orbit coupling at the interface of the magnetic layer and the heavy metal substrate combined with the broken inversion symmetry at interfaces [2]. The resulting spin states are spin spirals and skyrmions with a unique rotational sense enforced by the chiral DMI [3-5]. In general, spin spirals are periodic magnetic textures of helical, cycloidal, or conical nature, whereas magnetic skyrmions are particle-like, topologically distinct states which are either of cycloidal or helical nature. While they were theoretically predicted in the nineties [6], the first experimental evidences for the existence of skyrmions were reported only a few years ago [7-9]. Since then, skyrmions have attracted significant interest due to their potential applications in future spintronic devices [10-13].

Among the non-centrosymmetric bulk crystals that exhibit spin spirals and magnetic skyrmions are MnSi [7, 14-17], Fe$_x$Co$_x$Si [8, 18], FeGe [9], and Cu$_2$OSeO$_3$ [19]. These types of materials typically show helical order as magnetic ground state and the application of magnetic fields leads to spin spiral propagation reorientations [20-23], conical phases, and skyrmions. Recently, it was found that thin films of MnSi exhibit a skyrmion phase in a wider range of magnetic fields and temperatures compared to bulk crystals [15, 24]. In the low film thickness limit, i.e. ultrathin films of one or few atomic layers, the DMI is interface-induced [2-5]. Skyrmions in such ultrathin magnetic films are, in contrast to bulk non-centrosymmetric crystals, confined by the effectively two-dimensional structure. As a consequence, it is interesting to study the response of such interface-driven skyrmions to in-plane and canted magnetic fields, in contrast to skyrmions in bulk crystals which can always orientate along the externally applied field.

Symmetry breaking of spin spirals and skyrmions by in-plane and canted magnetic fields

In thin films of non-centrosymmetric materials magnetic anisotropy can pin magnetic textures to a particular direction relative to the crystal, allowing the distortion of spin spirals and skyrmions by canted magnetic fields [25]. Lin *et al.* simulated within a classical spin model the distortion of skyrmions due to canted fields which leads to an asymmetric shape of skyrmions [26]. A corresponding effect has been observed for chiral domain walls, leading to an asymmetric domain wall movement in canted fields [27]. These investigations already indicate that it is of fundamental interest to understand the influence of canted magnetic fields on the static and dynamic properties of magnetic skyrmions.

Here, we investigate the influence of in-plane and canted magnetic fields on the spin structure of atomic bilayer islands of palladium and iron on Ir(111) by using scanning tunnelling microscopy at low temperatures. We show that the propagation direction of the spin spiral is determined by the coupling to the islands' boundaries and thus by their shape. The application of in-plane fields along the spin spiral's propagation direction distorts the spin spiral, thereby revealing its cycloidal nature. The symmetry breaking by the in-plane fields allows, depending on the island's shape, a reorientation of the propagation direction of spin spirals which becomes perpendicular to the magnetic in-plane field. Furthermore, we prove that the skyrmions of this system are cycloidal and have the theoretically predicted anticlockwise sense of magnetisation rotation [28, 29] by applying canted fields causing a symmetry breaking for the previously rotational-symmetric skyrmions [30].

## 2. Spin spirals in Pd-Fe bilayer islands and their response to out-of-plane magnetic fields

Bilayer islands of Pd-Fe on Ir(111) were prepared similar to previous studies [5,30]. They show a spin spiral ground state in zero external field at low temperatures due to the competition of Heisenberg exchange and Dzyaloshinskii-Moriya (DM) interactions [5, 28, 29], as can be revealed by high-resolution spin-polarised scanning tunnelling microscopy (SP-STM) [31], see Fig. 1. While Fe grows in fcc stacking at the lower part of the Ir(111) atomic step edges, the Pd grows as either hcp or fcc stacked islands on top of the atomic Fe layer or at the lower step edge of the extended Fe film. Here, we focus on fcc stacked Pd islands on top of fcc grown Fe layers. The orientation of the hexagonal atomic lattice of the Pd-Fe bilayer, which is pseudomorphically grown on the Ir(111) substrate, can be derived from the straight edges of the Pd islands on the Fe monolayer, see Fig. 1a. Since SP-STM is sensitive to the projection of the local sample magnetisation onto the quantisation axis given by the SP-STM tip's magnetisation direction, the observed stripes on top of the topographically flat island correspond to the wavefronts of the spin spiral. Obviously, these stripes are not strictly parallel across the island but exhibit bends and even branches. The bends in the spin spirals indicate that there is no strong coupling to the hexagonal crystal lattice. Indeed, a comparison of the orientation of the hexagonal atomic lattice and the propagation direction of the spin spiral for an area in the interior of the island (as marked in Fig. 1a) shows that the spin spiral does not align with a high symmetry direction of the hexagonal atomic lattice. A closer inspection of the spin spiral's propagation vectors $k_{SS}$ reveals that the spin spiral prefers to propagate along the island's border as can be seen in Fig. 1a.

Symmetry breaking of spin spirals and skyrmions by in-plane and canted magnetic fields

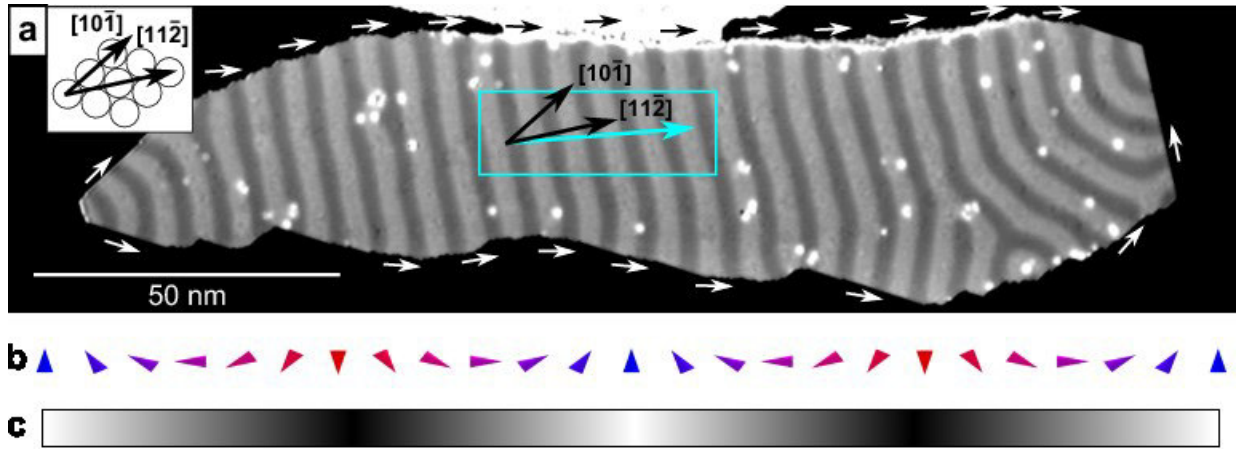

**Figure 1:** Spin structure of a Pd-Fe bilayer island on Ir(111). (a) Constant-current SP-STM map of a Pd-Fe bilayer island on Ir(111) exhibiting a spin spiral state (measurement parameters: $T = 8$ K, $V = 50$ mV, $I = 0.2$ nA). An out-of-plane magnetised bulk Cr tip was used for magnetic contrast imaging. The inset shows the orientation of the hexagonal atomic lattice with its crystallographic directions. The white arrows indicate the local spin spiral propagation direction at the island's border. The blue arrow displays the propagation direction within the area indicated by the blue rectangle. (b) Side-view schematics of a cycloidal spin spiral and (c) schematic top-view of the expected SP-STM contrast obtained with an out-of-plane sensitive probe tip.

In order to obtain deeper insight into the mechanisms of the arrangement of the spin-spirals with respect to the island's borders, we performed calculations of the equilibrium energy of a spin-spiral state on a triangular lattice. For simplicity, we take an effective nearest neighbour exchange interaction and DM interaction of strengths $J$ and $D$ per atom into consideration with a $D/J$ ratio close to the experimental system [32]. We find that a reorientation of $k_{SS}$ from the high symmetry direction $[11\bar{2}]$ to the $[10\bar{1}]$ direction reduces the energy by $1.44 \cdot 10^{-5} J$ for each atom within an extended magnetic film. In contrast, we can show that the energy of an atom at the rim along a close-packed row of a magnetic film depends much stronger on the direction of the propagation vector of the local spin-spiral configuration. The orientation of $k_{SS}$ parallel to the rim reduces the energy by $9.2 \cdot 10^{-2} J$ for each atom at the rim compared to an alignment of $k_{SS}$ perpendicular to the rim. In both cases, the internal energy of the spin-spiral state is minimised by an alignment of $k_{SS}$ with a crystallographic direction in such a way that all bonds contribute to the reduction of both the exchange energy and the DM energy. The larger contribution of the rim is a result of its related symmetry breaking. A more detailed explanation including the calculations can be found in the supplement. As an example, the exact values can be calculated for the island shown in Fig. 1. We determined the number of atoms at the border by dividing the perimeter by the nearest-neighbour distance (2.715 Å) of the pseudomorphic film. In the same way, the absolute number of atoms in the island was estimated by dividing the island's area by the area of a hexagonal unit cell. Finally, we can estimate that a reorientation of the spin spiral state can reduce the energy by $3.2$ meV within the magnetic island of Fig. 1, while a reduction of $350$ meV may be achieved at its border. Thus we find that the influence of the border is about 100 times larger than that of the inner part for this particular island. This means that the propagation direction of the spin spiral in the island is dominated by its boundary. We suggest that the observed frustration, the bends and branches, of the spin spiral propagation directions originates from the influence of the differently orientated sections of the island's borders.

Besides SP-STM measurements, which are directly sensitive to the projection of the local sample's magnetisation onto the quantisation axis given by the magnetisation direction of the SP-STM tip, the

Symmetry breaking of spin spirals and skyrmions by in-plane and canted magnetic fields recently discovered non-collinear magnetoresistance (NCMR) effect [33] can be exploited to investigate changes in the non-collinearity of the sample magnetisation even with a non-magnetic STM tip. In general, NCMR leads to a change in the measured differential tunnelling conductance, d$I$/d$V$, for increasing local non-collinearity of the sample's magnetic moments. For the Pd-Fe bilayer system, the NCMR scales roughly with the nearest-neighbour-angle along $k_{SS}$, $\theta_{nn}$, and results in a decrease of the d$I$/d$V$ signal with $\theta_{nn}$. Thus, relative changes of the magnetisation directions can be investigated by spatially resolved NCMR mapping. The use of a non-magnetic probe tip in NCMR based studies of the response of a magnetic system to an external field avoids ambiguities in data interpretation, compared to SP-STM studies with ferromagnetic probe tips where a possible reorientation of the tip's magnetisation direction in an external magnetic field might occur.

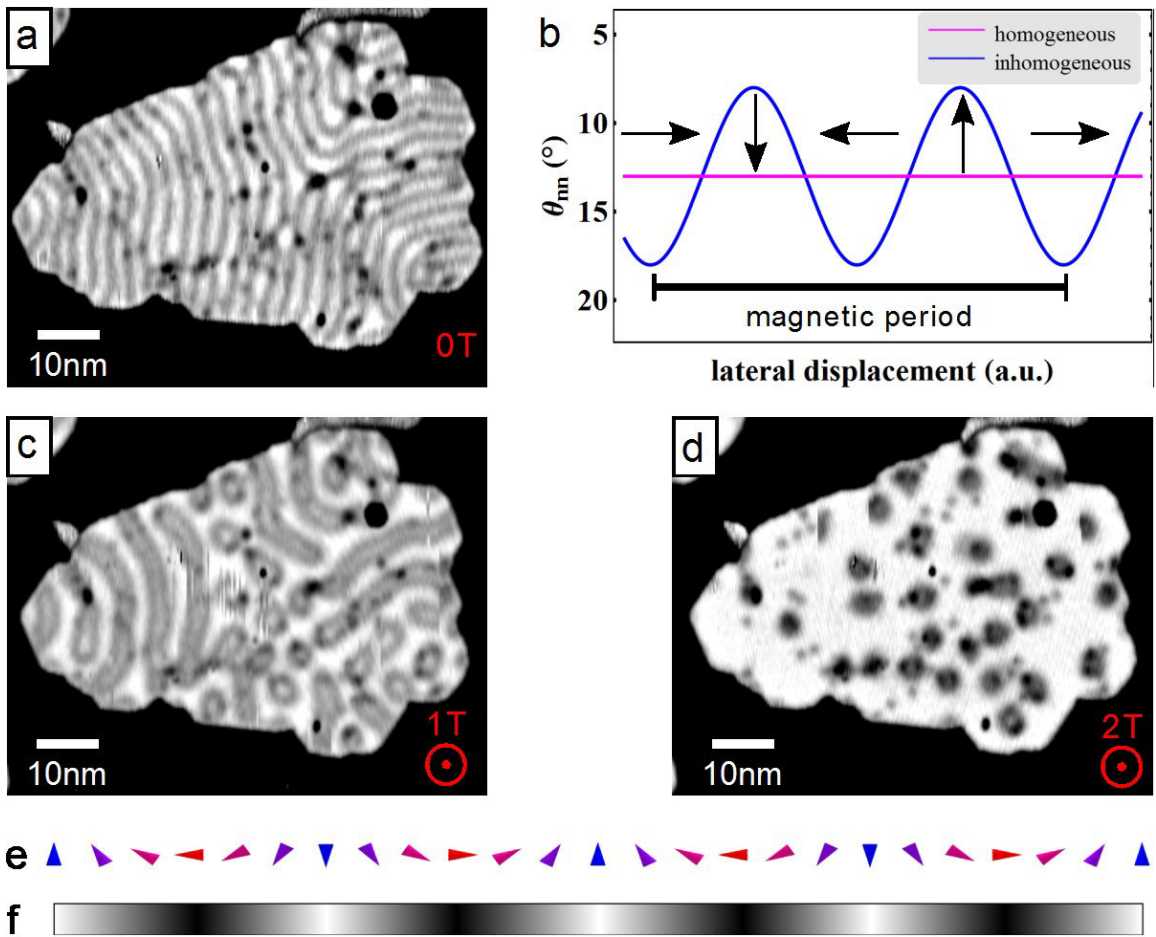

**Figure 2:** Spectroscopic d$I$/d$V$ maps of a Pd-Fe bilayer island on Ir(111) revealing its spin structure via NCMR contrast: (a) without an external magnetic field applied, and with a perpendicular field of (c) $B = 1\,\text{T}$ and (d) $B = 2\,\text{T}$ applied (measurement parameters: $T = 4.7\,\text{K}$, $V = 700\,\text{mV}$, $I = 2\,\text{nA}$). A non-magnetic STM tip was used. (b) Illustration of $\theta_{nn}$ along the propagation direction of a homogeneous and an inhomogeneous spin spiral with arrows indicating the corresponding magnetisation orientation. (e) Side-view schematics of a cycloidal spin spiral with (f) corresponding top-view of the NCMR contrast in zero field.

Symmetry breaking of spin spirals and skyrmions by in-plane and canted magnetic fields

Figure 2a shows the spin structure of a Pd-Fe bilayer island as a function of an externally applied magnetic field, as investigated with a non-magnetic STM tip making use of NCMR based imaging. In the Pd-Fe bilayer system, the NCMR contrast can be observed in a bias voltage interval of 600 mV to 800 mV; at other bias voltages this contrast vanishes. Since the Pd-Fe bilayer system is known to exhibit an out-of-plane easy axis [28-30], the spin spiral is inhomogeneous which means that the $\theta_{nn}$ of the magnetisation is oscillating between in-plane and out-of-plane parts of the spin spiral as illustrated in Fig. 2b. For this reason, the characteristic stripe pattern of the spin spiral can only be observed by NCMR imaging because of its inhomogeneity. A homogeneous spin spiral would display a constant d$I$/d$V$ signal. The dependence of the observed NCMR contrast on the change of $\theta_{nn}$ explains the halved period compared to the SP-STM contrast (see Fig. 2b,e,f). In Fig. 2c-d the response to an out-of-plane magnetic field is demonstrated. In the mixed phase at $B = 1$ T, spin spirals have already partially been transformed into skyrmions [5]. The part of the spin spiral with spins aligned parallel to the external field increases in width, while the skyrmions exhibit axially symmetric shapes [33]. For higher fields, at about $B = 2$ T, there are only single, pinned skyrmions left, see Fig. 2d.

### 3. Spin spirals in in-plane magnetic fields

Fig. 3a shows a Pd-Fe bilayer island in zero field while Fig. 3b,c show an island with an in-plane field applied which is either collinear or perpendicular to the spin spiral's propagation direction, respectively. While in the latter case no change in the appearance of the spin spiral is observed, a clear difference in the measured d$I$/d$V$ signal is revealed for those parts of the spin spiral with a local in-plane magnetisation direction for a field applied collinear to $\boldsymbol{k}_{SS}$. A detailed analysis of the observed changes based on line profiles is depicted in Fig. 3e, revealing clearly the alternating depth of the minima in the measured d$I$/d$V$. It is also evident that the difference of the conductance minima in applied field compared to the case of the field-free spin spiral is symmetric. As the change of the measured d$I$/d$V$ signal due to NCMR depends on $\theta_{nn}$, this finding demonstrates that the local non-collinearity of the spin spiral's in-plane parts changes depending on their alignment relative to the applied in-plane field. The fact that the changes in the measured d$I$/d$V$ only occur for fields applied along the $\boldsymbol{k}_{SS}$ is simultaneously an experimental proof for the cycloidal nature of the observed spin spiral. For a helical spin spiral, the in-plane parts of the local magnetisation are orientated perpendicular to the spin spiral's $\boldsymbol{k}_{SS}$. Therefore, such a spin spiral would exhibit the same change of contrast as the cycloidal spin spiral shows in Fig. 2b,c, however, for a perpendicular orientation of the in-plane field. In this way, we can also explain the result for a perpendicular in-plane field as seen in Fig. 3c. Here, we expect that the cycloidal spin spiral changes into a transversal-conical phase in order to arrange with the external field. In this case, the $\theta_{nn}$ are expected to decrease with increasing in-plane field. Micromagnetic simulations using OOMMF [34] and parameters obtained from experiments on hcp-stacked Pd-Fe bilayer islands [30] corroborate our explanation for the contrast changes as can be seen by the periodic change of $\theta_{nn}$ relative to the case without field, shown in Fig. 3d. The simulation has also been conducted for the case of an in-plane field perpendicular to the $\boldsymbol{k}_{SS}$ which reveals a significantly smaller decrease in $\theta_{nn}$ compared to the collinear field orientation. The strongest change in $\theta_{nn}$ occurs for the in-plane parts of the spin spiral. The simulation shows a deviation from the out-of-plane orientation towards the field direction of roughly 11 ° and 8 ° for the in-plane and the out-of-plane part of the spin spiral, respectively. Thus we expect the difference in d$I$/d$V$ between in-plane and out-of plane parts to decrease due to a perpendicular in-plane field which is, however, difficult to observe experimentally.

Symmetry breaking of spin spirals and skyrmions by in-plane and canted magnetic fields

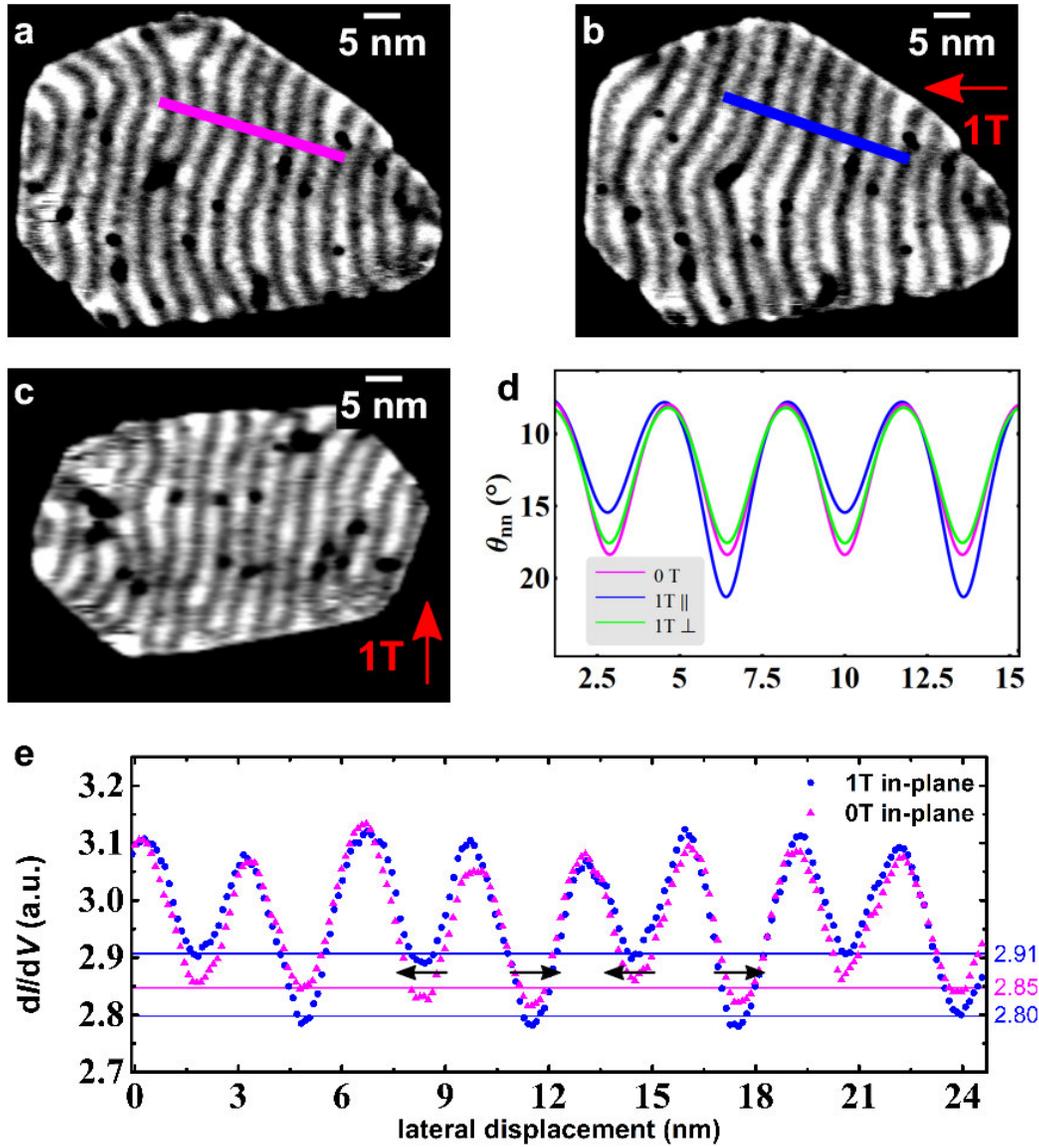

**Figure 3:** Spectroscopic d$I$/d$V$ maps of Pd-Fe bilayer islands on Ir(111) exhibiting a spin spiral (a) without an applied magnetic field and (b,c) in an in-plane field of $B = 1$ T as indicated. (d) Nearest-neighbour-angles along the propagation direction of a spin spiral simulated by OOMMF [34]. (e) Averaged line profiles of the areas marked in (a) and (b) with horizontal lines indicating the mean values for the minima. The black arrows illustrate the orientation of the magnetic moments derived from the change in the d$I$/d$V$ signal of (b).

Figure 4a shows several Pd-Fe bilayer islands at 4.7 K and in in-plane fields as indicated. At this temperature the application of in-plane fields only leads to the above mentioned distortions of the spin spiral. After warming the sample up to 30 K and cooling it down again, while the field is applied, the $k_{SS}$ of some of the islands reorientate to become perpendicular to the field (see Fig. 4b). This demonstrates that the application of in-plane fields leaves the system in a metastable state as it lacks the energy to overcome the barriers to lower energy states. The system is now in a field-cooled state while the former orientation of the spin spiral was related to a metastable state. In Fig. 4c-d several islands can be seen that were subsequently field-cooled in in-plane fields perpendicular to each other. On the highlighted island the largest part of the spin spiral switches its $k_{SS}$ to become perpendicular to the field. The other islands show also varying degrees of change in the $k_{SS}$ of the spin spirals. Thus a reorientation of the spin spiral's propagation direction due to an in-plane field is possible, but competes with the strong coupling to the island's rim. These results show that a spin spiral propagation perpendicular to a magnetic in-plane field is energetically preferred.

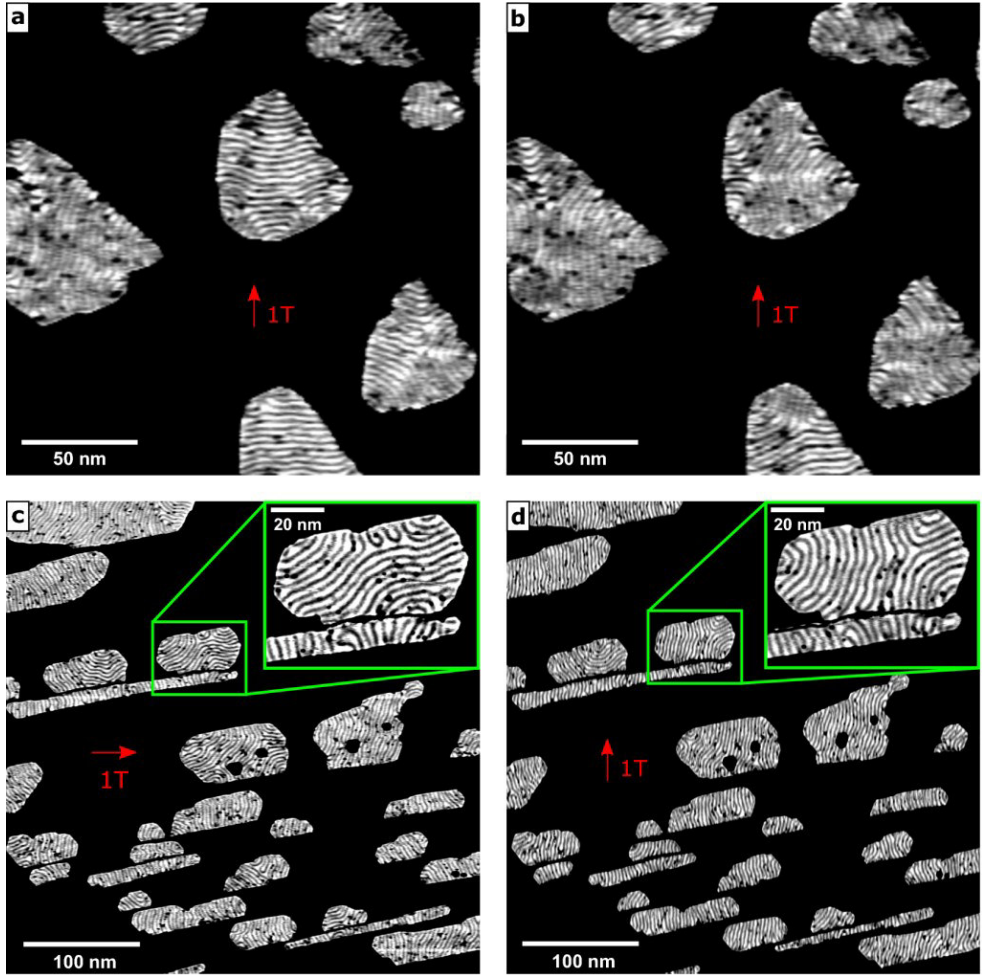

**Figure 4:** Spectroscopic d$I$/d$V$ maps of Pd-Fe bilayer islands obtained with a non-magnetic STM tip. While the map in (a) shows spin spirals in an in-plane field as indicated, the maps of (b-d) display the field-cooled (from 30 K to 4.7 K) states for in-plane fields as indicated. The maps of the highlighted areas are separate scans where small changes in the propagation direction of the spin spirals can be recognised.

## 4. Skyrmions in canted magnetic fields

Application of out-of-plane magnetic fields leads to a phase transition from the spin spiral state to the skyrmionic state [5], as shown in Fig. 2. For the out-of-plane fields used here the in-plane part of the skyrmion shows the highest non-collinearity and thus the lowest d$I$/d$V$ signal [33] (compare to illustration in Fig. 5e). Figures 5a-b show numerous skyrmions in canted fields. The STM measurement parameters that are necessary to observe the NCMR contrast can lead to annihilation and creation of skyrmions on the time-scale of the STM scan, thereby causing sudden jumps, i.e. line noise, in the STM images. While in Fig.

5a the skyrmions have a tendency to exhibit a lower d$I$/d$V$ on their left side, the skyrmions in the inverted in-plane field show likewise a tendency of a reduced d$I$/d$V$ signal on their right side (see Fig. 5b). A particularly stable skyrmion is marked and examined by high-resolution maps (Fig. 5c,d) together with the corresponding line profiles along the direction of the applied in-plane field (Fig. 5f). The observed change in d$I$/d$V$ is similar to the one for the spin spirals: again, the d$I$/d$V$ changes according to the orientation of the in-plane parts of the spin structure relative to the direction of the applied canted magnetic field. Since in NCMR contrast images the change in d$I$/d$V$ signal depends roughly on the change of $\theta_{nn}$ we conclude that the $\theta_{nn}$ in the in-plane parts of the spin structure, that are collinear to the in-plane part of the canted field, change asymmetrically. The previously rotational symmetric skyrmion [30] experiences a symmetry breaking and is transformed into a non-circular skyrmion. The canted field reduces the rotational symmetry to a mirror plane symmetry along the direction of the in-plane field component. However, the exact shape of the skyrmion strongly depends on the vicinity to other skyrmions or defects. Especially, the measured d$I$/d$V$ in the area between adjacent skyrmions, which exhibits a smaller degree of non-collinearity in the orientation of the magnetic moments, depends on the distance between the skyrmions or skyrmions and defects in the Pd-Fe bilayer. This explains the different d$I$/d$V$ signals at the rims of the skyrmions that can be observed in the line-profiles of Fig. 5f. The response to the canted magnetic field confirms that the observed skyrmions are cycloidal. For a helical skyrmion the asymmetry in the in-plane parts would show up on the axis perpendicular to the applied field direction. The data obtained also allows a determination of the sense of rotation as enforced by the DMI. The observed asymmetry reveals in which direction the in-plane parts of the local magnetisation, that are collinear to the applied field, are pointed and - in combination with the orientation of the applied field's out-of-plane component - we can derive the perpendicular orientation of the magnetic moments in the center of the skyrmion and its surrounding (compare Fig. 5e). The spatial distribution of the magnetic moments (as illustrated by black arrows in Fig. 5f) is found to exhibit an clockwise rotation, and thereby our experimental results confirm the theoretical predictions by B. Dupé *et al.* [28] and E. Simon *et al.* [29].

Symmetry breaking of spin spirals and skyrmions by in-plane and canted magnetic fields

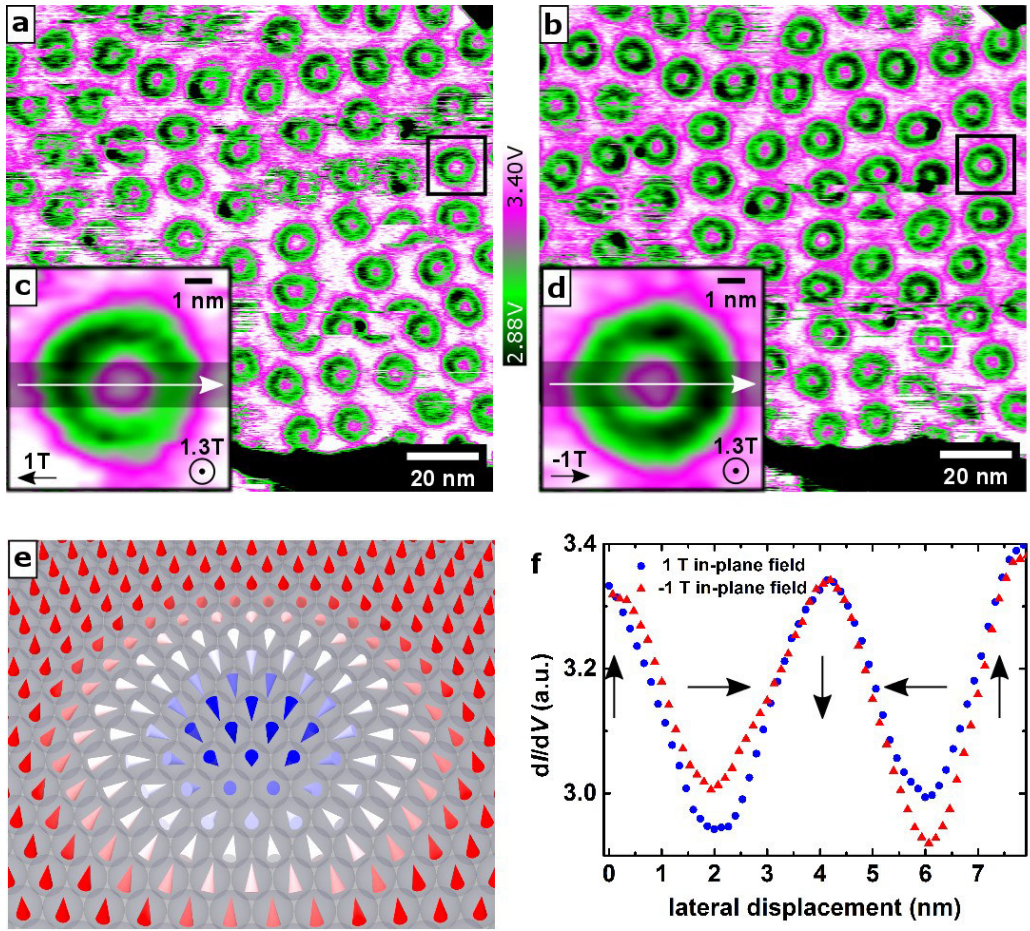

**Figure 5:** Spectroscopic d$I$/d$V$ maps of a Pd-Fe bilayer island in a perpendicular field of $B$ = 1.3 T and an in-plane field of (a) $B$ = 1 T and (b) $B$ = -1 T. (c) and (d) show Gauss-filtered d$I$/d$V$ maps of single skyrmions extracted from (a) and (b) (marked by black squares) in magnetic fields as indicated. (e) Illustration of a cycloidal skyrmion with cones representing the magnetic moments. Averaged line profiles taken from (c) and (d) are shown in (f). The shaded areas in (c) and (d) show the regions used for the averaged line profiles (f).

## 5. Conclusion

We have shown that the spin spirals in Pd-Fe bilayer islands prefer to propagate along the islands' borders. A calculation corroborated the experimental results showing that the coupling of the spin spiral to the border is by orders of magnitude stronger than the coupling to a particular symmetry direction of the hexagonal atomic lattice. In-plane magnetic fields change the nearest-neighbour-angles between magnetic moments of the spin spiral's in-plane parts if applied collinear to the spin spiral's propagation direction, thereby providing an experimental proof that the spin spiral is cycloidal. In contrast, for a perpendicular orientation of the in-plane field relative to the spin spiral's propagation direction, our simulations suggest a distortion towards a transversal-conical spin spiral. Field-cooled samples in differently oriented in-plane fields lead, depending on the island's shape, to a reorientation of the spin spiral propagation direction. On the other hand, the application of magnetic fields at 4.7 K only results in distortions of the spin spiral. A canted field induces an asymmetry in the skyrmion's shape along the field direction, thereby breaking its rotational symmetry. For skyrmions the induced asymmetry not only reveals their cycloidal nature, but additionally allows the determination of the sense of magnetisation rotation.

Symmetry breaking of spin spirals and skyrmions by in-plane and canted magnetic fields

**Acknowledgements**

Financial support from the Deutsche Forschungsgemeinschaft (SFB 668-A8 and GrK 1286), the European Union (FET-Open MAGicSky No. 665095), and the Hamburgische Stiftung für Wissenschaften, Entwicklung und Kultur Helmut und Hannelore Greve is gratefully acknowledged. We thank C. Hanneken, B. Dupé and E. Vedmedenko for useful discussions.

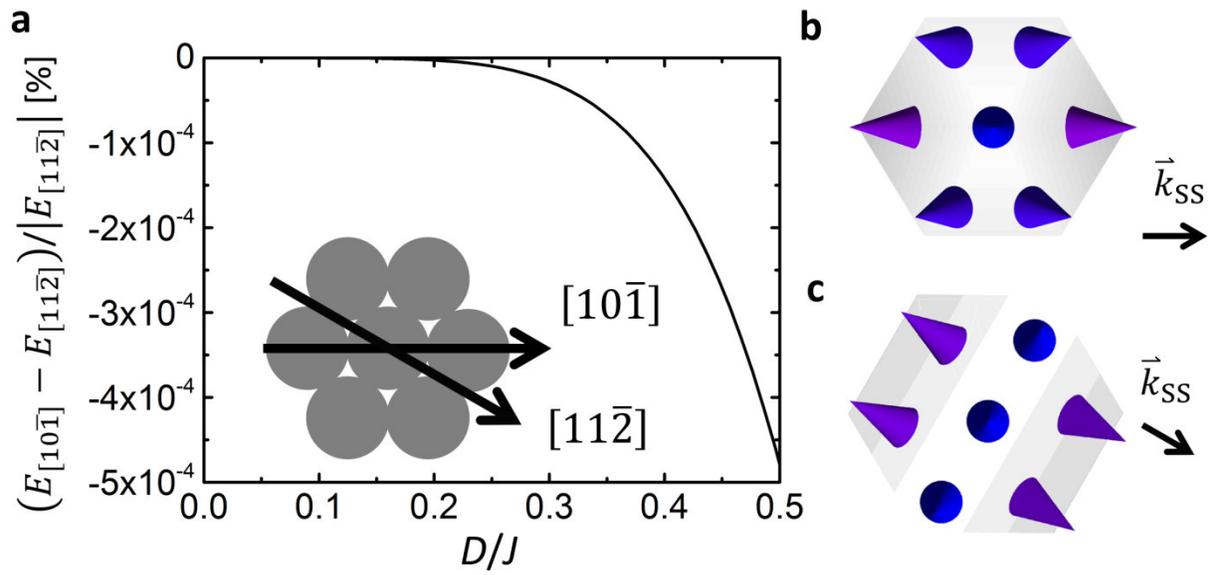

Supplementary Figure 1 | **Spin-spirals on a triangular lattice**. (a) Energy difference between the states where $\mathbf{k}_{ss}$ is aligned parallel to the $[10\bar{1}]$ and the $[11\bar{2}]$ directions. (b), (c) Cones illustrating the directions of the atomic magnetic moments in spin-spirals with $\mathbf{k}_{ss}$ parallel to the crystallographic $[10\bar{1}]$ and $[11\bar{2}]$ directions.

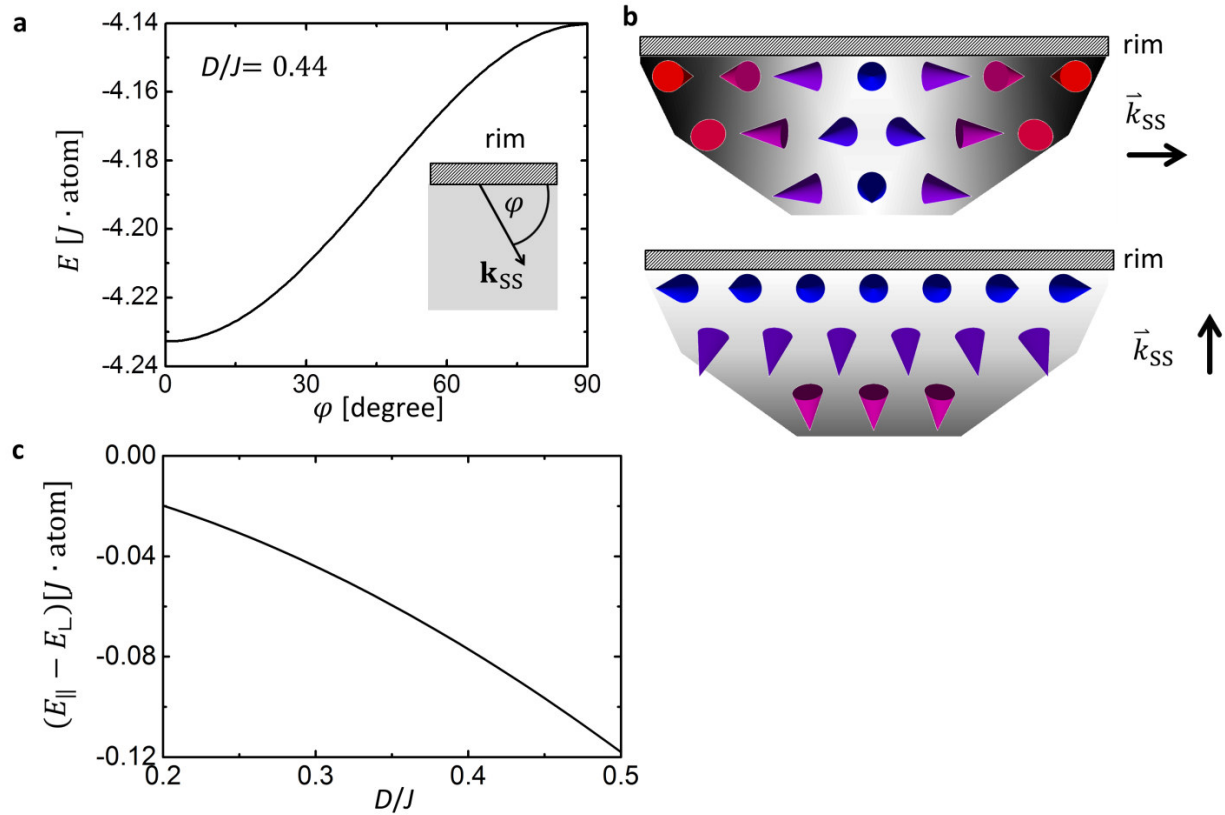

Supplementary Figure 2 | **Spin-spirals on a triangular lattice - effects of a boundary**.
(a) Energy of an atom at a close packed border of an island as a function of the angle between $\mathbf{k}_{ss}$ and the border. (b) Cones illustrating the directions of the atomic magnetic moments in spin-spirals at the rim of a magnetic island for $\mathbf{k}_{ss}$ parallel and perpendicular to the rim. (c) Energy difference per border atom between parallel and perpendicular alignment of $\mathbf{k}_{ss}$ with respect to the border as a function of $D/J$.

**Supplementary Note 1: Spin-spirals on a discrete lattice**

In the following, we describe the influence of an underlying discrete triangular crystal lattice on the alignment of a spin-spiral state. We consider an ultrathin magnetic layer and describe it with the Hamiltonian

$$H = -J \sum_{<i,j>} S_i S_j - \sum_{<i,j>} D_{i,j} \cdot (S_i \times S_j)$$

which takes an effective nearest-neighbour exchange interaction and nearest-neighbour DMI of strengths $J$ and $D$ per atom into consideration. We assume the DM-vectors $D_{i,j}$ to be perpendicular to the connection lines of two atomic sites and to lie within the plane of the magnetic film[1]. $S_i$ is considered to be of unit length and can take any orientation in three-dimensional space. The interactions are known to lead to spin-spiral states at zero external magnetic fields and at low temperatures.

We consider the energies $E_{[11\bar{2}]}$ and $E_{[10\bar{1}]}$ of spin-spirals for which the corresponding wave vectors $k_{iss}$ are aligned with either of the two high symmetry directions $[11\bar{2}]$ and $[10\bar{1}]$. The total energy of a homogenous spin-spiral state can be calculated as a multiple of the energy of a single atom which is

$$E_{[11\bar{2}]}(\theta) = -2J - 4J\cos(\theta) - 2\sqrt{3}D\sin(\theta)$$
$$E_{[10\bar{1}]}(\theta) = -2J\cos(2\theta) - 2D\sin(2\theta) - 4J\cos(\theta) - 2D\sin(\theta)$$

where $\theta$ is the angle between the magnetisation directions of neighbouring ferromagnetic atomic rows. The energy $E_{[11\bar{2}]}$ can be minimised analytically with respect to $\theta$ providing

$$\tan\left(\theta_{[11\bar{2}]}^{min}\right) = \frac{\sqrt{3}}{2}\frac{D}{J}$$

while $E_{[10\bar{1}]}$ needs to be minimised numerically. We find that the deviation between the two energies increases with $D/J$ as shown in Fig. 1a and that $k_{ss}$ aligned with the $[10\bar{1}]$ direction is the energetically more favorable state. The energetic preference of this direction is due to the fact that all bonds reduce both the DM-energy and the exchange energy which is not true for an alignment of $k_{ss}$ parallel to the $[11\bar{2}]$ direction (Fig. 1b,c).

Moreover, we investigate the energy of a magnetic atom at the rim of an island and find that the energy is most favourable when $k_{ss}$ is parallel to the border and most unfavourable when it is perpendicular to the border which is shown exemplarily for $D/J = 0.44$ in Fig. 2a. The reason for the parallel alignment is once more that all bonds reduce both the exchange energy and the DM-energy for this spin configuration (Fig. 2b). The energy difference between a parallel and perpendicular alignment is given in Fig. 2c as a function of $D/J$.

We combine the findings from beforehand with the effective values for $D$ and $J$ derived from density functional theory calculations[2] for fcc Pd/Fe/Ir(111) in order to estimate the absolute values of the energy contributions of the rim and the inner part of an experimentally investigated island.